\documentclass[10pt,a4paper,twoside]{article}
\usepackage{epsfig}
\usepackage{baltlat6}
\usepackage{array}
\usepackage{here}
\begin{document}
\ \
\vspace{0.5mm}
\setcounter{page}{223}

\titlehead{Baltic Astronomy, vol. 24, 223--230, 2015}

\titleb {STOCHASTIC 2-D GALAXY DISK EVOLUTION MODELS. \\ RESOLVED STELLAR POPULATIONS IN THE GALAXY M33}

\begin{authorl}
\authorb{T. Mineikis}{1,2} and
\authorb{V. Vansevi\v{c}ius}{1,2}
\end{authorl}

\begin{addressl}
\addressb{1}{Vilnius University Observatory,
\v{C}iurlionio 29, Vilnius LT-03100, Lithuania; 
vladas.vansevicius@ff.vu.lt}
\addressb{2}{Center for Physical Sciences and Technology,
 Savanori\c{u} 231, Vilnius LT-02300, Lithuania;
tadas.mineikis@ftmc.lt}
\end{addressl}

\submitb{Received: 2015 June 04; accepted: 2015 July 14}

\begin{summary}
We improved the stochastic 2-D galaxy disk models (Mineikis
\& Vansevi\v{c}ius 2014a) by introducing enriched gas outflows from galaxies and
synthetic color-magnitude diagrams of stellar populations. To test the models,
we use the HST/ACS stellar photometry data in four fields located along the
major axis of the galaxy M33 (Williams et al. 2009) and demonstrate the
potential of the models to derive 2-D star formation histories in the resolved
disk galaxies.
\end{summary}


\resthead{Resolved stellar populations in M33}
{T. Mineikis, V. Vansevi\v{c}ius}

\sectionb{1}{INTRODUCTION}

One of the most informative sources to study the evolution of galaxy disks is
a resolved stellar photometry. Accurate photometric data supported by realistic
galaxy evolution models allow one to derive the most important galaxy parameters
-- star formation (SF) and metal enrichment histories, see, e.g., Aparicio \& Hidalgo
(2009). Moreover, the possibility of determining 2-D star formation (SF) histories
over wide areas offers perhaps the best opportunity to peer into large scale (beyond
giant molecular clouds) SF processes in the galaxy disks.

Recent studies based on wide field and deep stellar photometry data in nearby
galaxies revealed spatially resolved SF histories, see, e.g., Harris \& Zaritsky (2009),
Rubele et al. (2015), Lewis et al. (2015). Such studies, focussing on the large scale
SF in galaxies, could lead to better understanding of the SF processes in general.
For this purpose the 2-D stochastic galaxy disk evolution models (Mineikis \&
Vansevi{\v c}ius 2014a) could serve as an alternative to the widely employed, however,
computationally expensive hydrodynamical models.

In this paper, we extend further our study of the galaxy M33 (Mineikis \&
Vansevi{\v c}ius 2014a) using improved 2-D stochastic galaxy disk models. The models
are improved in two aspects: outflows of the enriched gas are considered and
capabilities to generate synthetic stellar photometry catalogs are implemented.
These new features enhanced the models and made them useful for the analysis of
spatially resolved SF histories in disk galaxies.

\sectionb{2}{THE MODELS}

In this section we only briefly sketch our 2-D galaxy disk models, which are
presented in Mineikis \& Vansevi{\v c}ius (2014a) and compared with 1-D cases in
Mineikis \& Vansevi{\v c}ius (2014b). Below we discuss in more detail only new features
implemented in the models recently: a prescription of gas outflows from galaxies
and a method of synthetic stellar photometry data generation.

\subsectionb{2.1}{2-D galaxy disk models}

The 2-D galaxy disk models are composed of concentric rings subdivided into
cells of equal area. The disk mass is building up by a gradual gas accretion from
the reservoir where gas resides initially. The accretion rate follows a mean dark
matter accretion rate as reported by Fakhouri, Ma \& Boylan-Kolchin (2010). The
SF is simulated by stochastic SF events in the model cells. The main model
parameters determining stochastic SF are the probability of triggered SF, $P_{\rm T}$, and
the SF efficiency (SFE). $P_{\rm T}$ controls the intensity of propagating SF. SFE depends
on the average gas density in bursting cells, $\Sigma_{\rm G}$, and two parameters, $\epsilon$ and $\alpha$:

\begin{equation}
{\rm SFE} = \epsilon\cdot \left( \frac{\Sigma_{\rm G}}{10\,
M_{\odot}/{\rm pc^2}} \right)^{\alpha}.
\end{equation}

\subsectionb{2.2}{Gas outflows}

In the previous study (Mineikis \& Vansevi{\v c}ius 2014a) we introduced gas flows
between the cells, occurring due to SF events in the neighboring cells. In this
study we consider, in addition, the case when a fraction of gas residing in a cell is
ejected during the SF event and leaves the galaxy.

We implemented a model of expanding super-bubbles in stratified atmospheres
following Baumgartner \& Breitschwerdt (2013). We assume that, after the SF
event in a cell, all newly formed supernovae explode in the center of the cell,
which is located exactly at the mid-plane of the disk. We derive the mid-plane gas
density assuming the conditions of hydrostatic equilibrium and a constant, along
disk radius, scale height, $h_{z}=100$\,pc. These approximations make the speed of
the expanding super-bubble in the $z$ direction (perpendicular to the galaxy disk)
dependant only on the gas density in the cell and the mass of newly formed stars
(luminosity of the last stellar population).

Baumgartner \& Breitschwerdt (2013) derived the conditions for disk gas out-
flow for the Milky Way by assuming that the expansion velocity of the top of
super-bubble reaches a critical value, $v_{z,{\rm c}}=20$\,km\,s$^{-1}$, before the acceleration
(due to the exponential drop in gas density) in the $z$ direction begins. We applied
the same critical value to our models of the galaxy M33.

In our models we assume that, once the outflow condition is fulfilled for the SF
event, the cell loses gas instantly. We also assume that the gas lost from the cell
is composed of the ejecta produced during the last 10\,Myr from dying stars of all
populations residing in it. Therefore, the mass of gas lost from the cell $i$, $m_{{\rm OUT}, i}$,
after the SF event is:

\begin{equation}
m_{{\rm OUT}, i} = \left\{ 
  \begin{array}{l l}
     0,      & n_{\rm SN} \le f(\Sigma_{{\rm G}, i}) \\
     \eta \cdot F_{i},  & n_{\rm SN} > f(\Sigma_{{\rm G}, i}), \\
  \end{array} \right.
\end{equation}

\noindent where $n_{\rm SN}$ is an average number of supernovae expected from the SF event, $f(\Sigma_{{\rm G}, i})$ denotes the Baumgartner \& Breitschwerdt (2013) condition for the outflow, $F_{i}$ is
a fraction of gas returned during the last 10\,Myr from all populations residing in
the cell, and $\eta$ is a free parameter controlling the fraction of the lost gas. If the
SF event produces $n_{\rm SN} \le 2$, the outflow from the cell is suppressed.

\subsectionb{2.3}{CMDs generation}

For each disk model cell we keep a complete star formation history in a matrix
$S(t, Z)$ with age $t$ and metallicity $Z$ dimensions. The step in age dimension, $\delta t$, is
equal to the model integration time step. The step in metallicity dimension varies
in the range 0.1$-$0.3\,dex. Each stellar population, corresponding to the particular
matrix element, at birth is generated by using a stochastically sampled initial
mass function (IMF). We assume that stellar population $j$, corresponding to the
matrix element $S(t_j, Z_j)$, has ages uniformly sampled within the time interval $\delta t$.
For each stellar population $j$, we find in the library the nearest by age isochrone
and interpolate the luminosity for each star of this population. However, we
do not interpolate stellar luminosities between metallicities. See Table\,1 for the
parameters used to generate CMDs.

\subsectionb{2.4}{Model parameters}

We use the same set-up of the model grid as in Mineikis \& Vansevi{\v c}ius (2014a).
Taking into account that gas outflows, implemented in the models, remove only
enriched gas, the strongest effect is expected on the evolution of galaxy's metallic-
ity. Since the enriched gas makes up a small fraction of the total mass of a galaxy,
its removal can have only a negligible effect on the galaxy's total mass budget.

For the present study we used three models possessing the following values
of parameters $P_{\rm T}$ and $\epsilon$ along the ``degeneracy valley'' (Mineikis \& Vansevi{\v c}ius
2014a): 0.3 and 1.0\% (the model A), 0.34 and 0.2\% (the model B), and 0.44 and
0.1\% (the model C), respectively. For simplicity, we kept the parameter $\alpha$ fixed at
the value $\alpha=2$ and generated three corresponding sets of models by varying the
outflow parameter $\eta$ in the range 0.2-0.9.

\begin{table}
\parbox[c]{124mm}{\baselineskip=0pt
{\smallbf\ \ Table\,1.}{The parameters used to generate synthetic CMDs.\lstrut}}
\begin{tabular}{l l l}     
\hline               
Parameter &  Description & Reference\\   
\hline                       
Isochrone library   	     & Parsec 1.2S & Bressan et al. (2012),\\
 & & Chen et al. (2014),\\
 & & Tang et al. (2014) \\
 Mass loss on RGB & $\eta_{\rm Reimers}=0.2$ & default value \\
Initial mass function        & corrected for binaries  & Kroupa (2002) \\
\hline                                   
\end{tabular}
\end{table}
\vskip5mm

\sectionb{3}{RESULTS}

A comparison of the updated stochastic 2-D galaxy disk models with the observational data for the galaxy M33 is presented in Fig.\,1, where the radial profiles of
gas surface density, surface brightness in the $i$ and $FUV$ passbands, and metallicity are shown. The observed radial profiles of surface brightness are de-projected
by adopting  54$^\circ$ for the inclination of the galaxy's disk. In the models, metallicity is converted to the oxygen abundance, with the solar metallicity, $Z_{\sun}=0.015$ (Bressan et al. 2012), and the solar oxygen abundance of 8.67 (Asplund et al. 2009).

\begin{figure}[!t]
\vbox{
\centerline{\psfig{figure=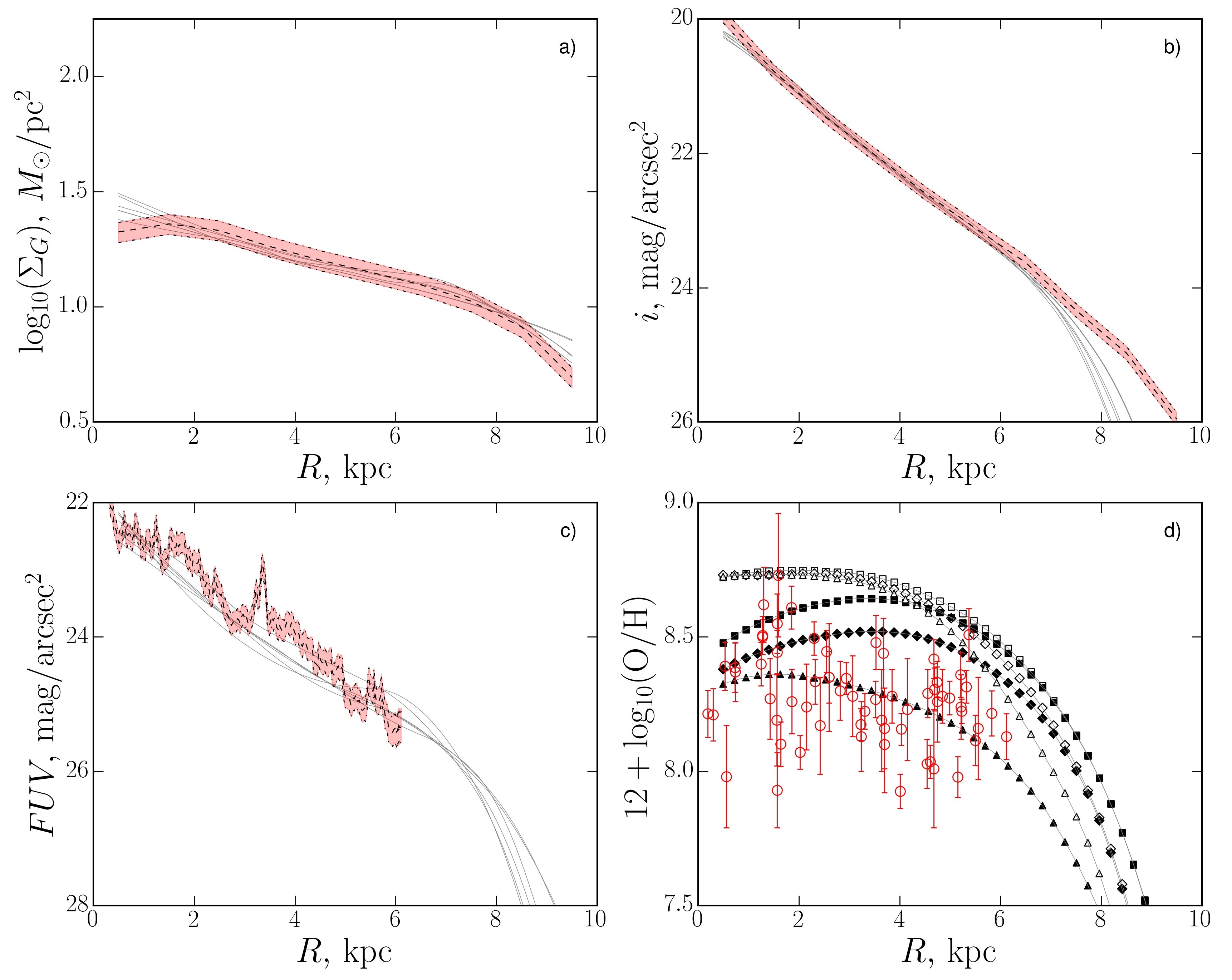,width=120mm,clip=}}
\vspace{1mm}
\captionb{1}
{Comparison of the observation data for the galaxy M33 with models: a)
radial profiles of the gas surface density obtained by co-adding H I (Corbelli \& Salucci
2000) and H$_2$ (Heyer et al. 2004) gas surface density profiles; b) radial profiles of the
surface brightness in the $i$ passband (Ferguson et al. 2007) deprojected to face-on and
corrected for internal extinction using the radial extinction profile (Munoz-Mateos et
al. 2007) and assuming the LMC type extinction law (Gordon et al. 2003); c) radial
profiles of the surface brightness in the GALEX $FUV$ passband (Munoz-Mateos et al.
2007) deprojected to face-on and corrected for internal extinction; d) radial profiles of
oxygen abundance of HII zones by Rosolowsky \& Simon (2008) (empty circles with error
bars) and model oxygen abundance (triangles -- $P_{\rm T}=0.3$ and $\epsilon=1\%$, diamonds -- $P_{\rm T}=0.34$ and $\epsilon=0.2\%$, squares -- $P_{\rm T}=0.44$ and $\epsilon=0.1\%$; open symbols correspond
to $\eta=0.2$, filled symbols to $\eta=0.9$).}}
\end{figure}

The new models produce ``degeneracy valley'' almost identical to that demonstrated in Mineikis \& Vansevi{\v c}ius (2014a). The only significant difference is seen
in the metallicity radial profiles, because the model galaxies lose different amount
of metal enriched gas due to different SF efficiency along the ``degeneracy valley''.
The models with the highest $\epsilon$ values experience more massive SF events which
are capable of producing outflows even in the outer disk regions. In contrast, the
models with the lowest $\epsilon$ values form less massive stellar populations, i.e., cells lose
a smaller amount of the enriched gas. The radial profiles of the enriched gas loss
also differ between the models. In the central parts of the galaxy, all models lose
a similar fraction of the enriched gas, however, in the outer parts the difference
becomes more pronounced. The models with highest values of $\epsilon$ lose significantly
more metals in the outer parts compared to the lowest $\epsilon$ models. Therefore, the differences in the metal enrichment history of the models make it possible to break
the parameter degeneracy.

\begin{figure}[!t]
\vbox{
\centerline{\psfig{figure=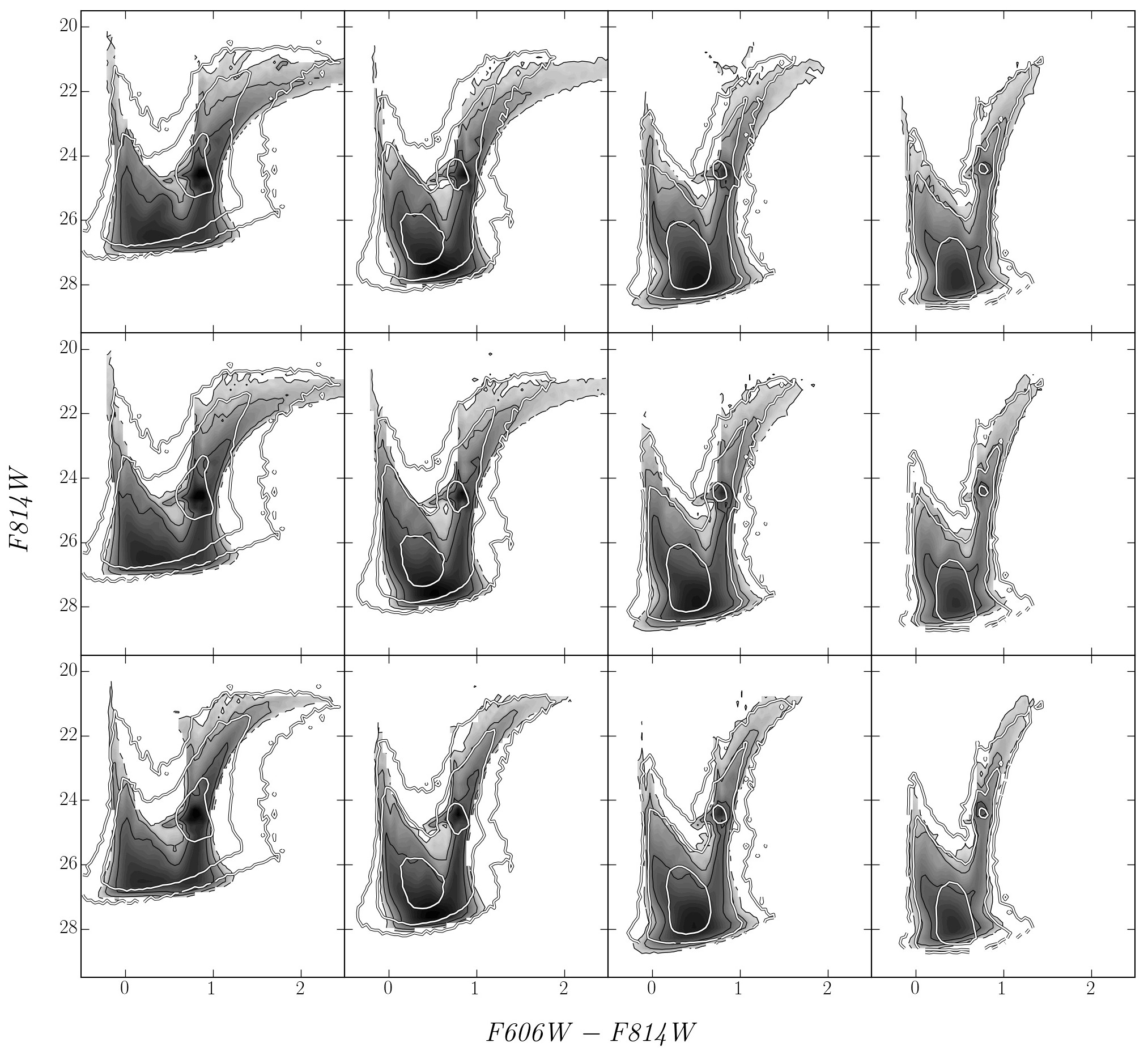,width=120mm,clip=}}
\vspace{1mm}
\captionb{2}
{HST/ACS WFC resolved stellar photometry data of the galaxy M33 (Williams
et al. 2009). The contour lines represent the observed CMD star density on the logarith-
mic scale (adjacent lines indicate density differing by 10) compared with the model A
CMDs plotted in grey (logarithmic scale). The rows of panels from top to bottom show
the model CMDs with $\eta=0.2$ (top row), $\eta=0.5$ (middle row), and $\eta=0.9$ (bottom
row). The columns of panels from right to left show CMDs sampled at radial distances
of 0.9, 2.4, 4.1, and 5.8\,kpc from the M33 center.}}
\end{figure}

However, to constrain galaxy SF parameters, the resolved stellar photometry is
necessary. For the CMD analysis we used the HST/ACS WFC stellar photometry
data in four fields within the galaxy M33, kindly provided by Benjamin Williams
(Williams et al. 2009).

The sizes of areas, used to sample models at different radial distances, are the
same as those of areas measured in Williams et al. (2009). To compare the model
and observed CMDs we adopt for M33 a distance of 840\,kpc, the same extinction
for all fields, $A_{V} = 0.3$, and apply to the model stars Gaussian errors calculated
using the relation between photometric errors and magnitude, which was derived
from the observed star catalog individually for each field. The extinction for the
HST/ACS bands was calculated according to Cardelli, Clayton \& Mathis (1989), with $R_{V} = 3.1$. Note, however, that in estimating the errors we do not take into
account the effects of crowded field photometry and differential extinction within
the galaxy M33.

\begin{figure}[!t]
\vbox{
\centerline{\psfig{figure=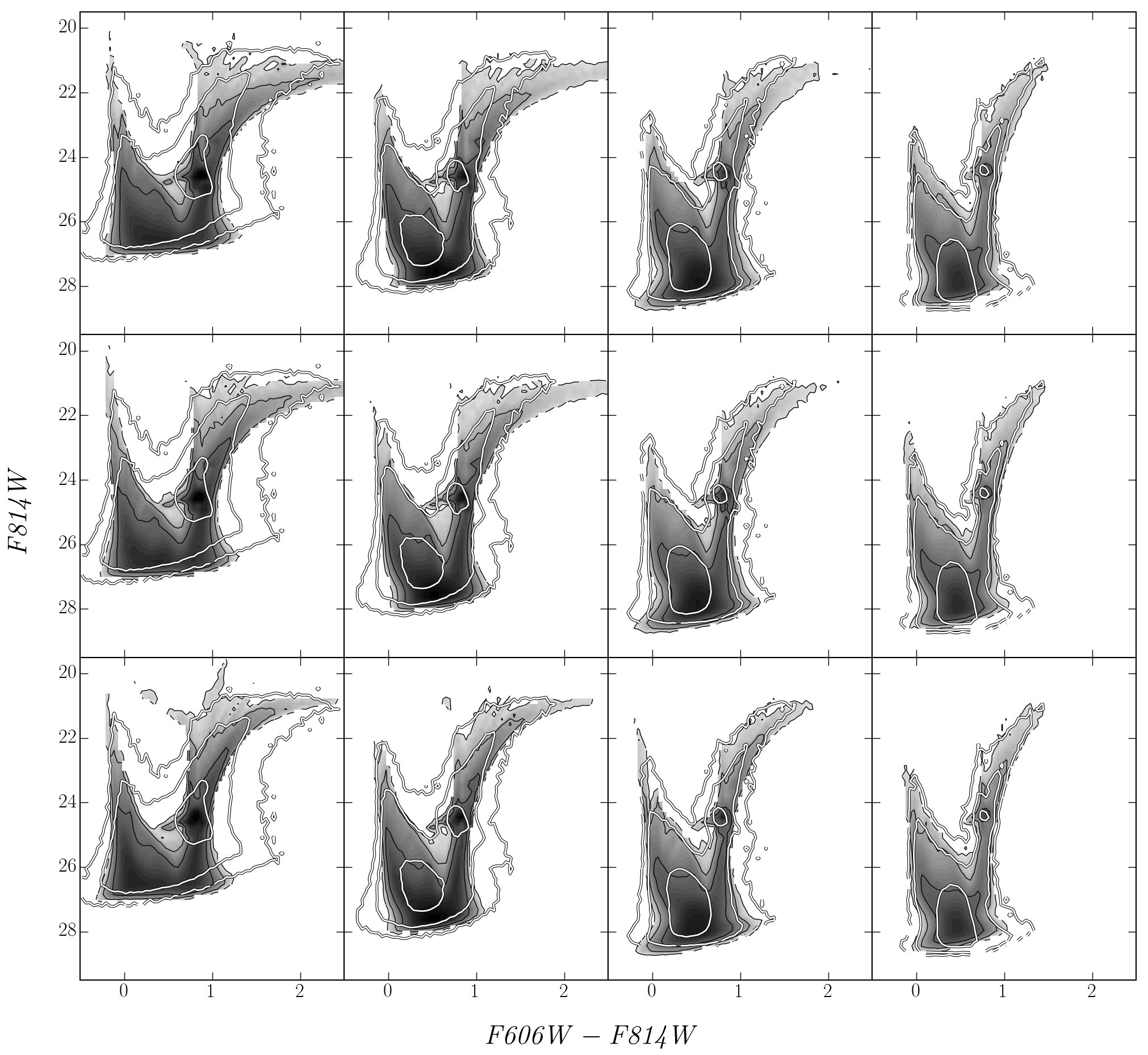,width=120mm,clip=}}
\vspace{1mm}
\captionb{3}
{The same as in Fig.\,2, but for the model B.}}
\end{figure}

In Fig.\,2 we plotted CMDs of the model A with different values of the outflow
parameter η, together with the observational data for M33. The model CMDs
with $\eta = 0.2, 0.5, 0.9$ are shown in the top, middle, and bottom rows of panels,
respectively. The model CMDs sampled at different distances from the galaxy
center, 0.9, 2.4, 4.1, and 5.8\,kpc, are shown in the columns of panels from the left
to the right, respectively. Fig.\,2 shows that CMDs of the model A are incompatible
with the M33 data. With $\eta = 0.2, 0.5$, only the CMDs in the outermost fields
match the observations. The inner fields display features of clearly too metallic
stellar populations in comparison with the observed data. The model CMDs with
the highest value of the outflow parameter, $\eta=0.9$, in the inner fields match the
observational data, however, in the outer fields, the models are clearly too metal
deficient.

In Fig.\,3 we compare, in a similar way, the observations with the model B. In
the cases of $\eta = 0.2, 0.5$ we see, again, the CMD features of too metallic stellar
populations in the inner fields. However, contrary to the model A, in the case of $\eta = 0.9$ (bottom row of panels), the CMDs of the model B match the observations
both in the inner and in the outer fields.

\begin{figure}[!t]
\vbox{
\centerline{\psfig{figure=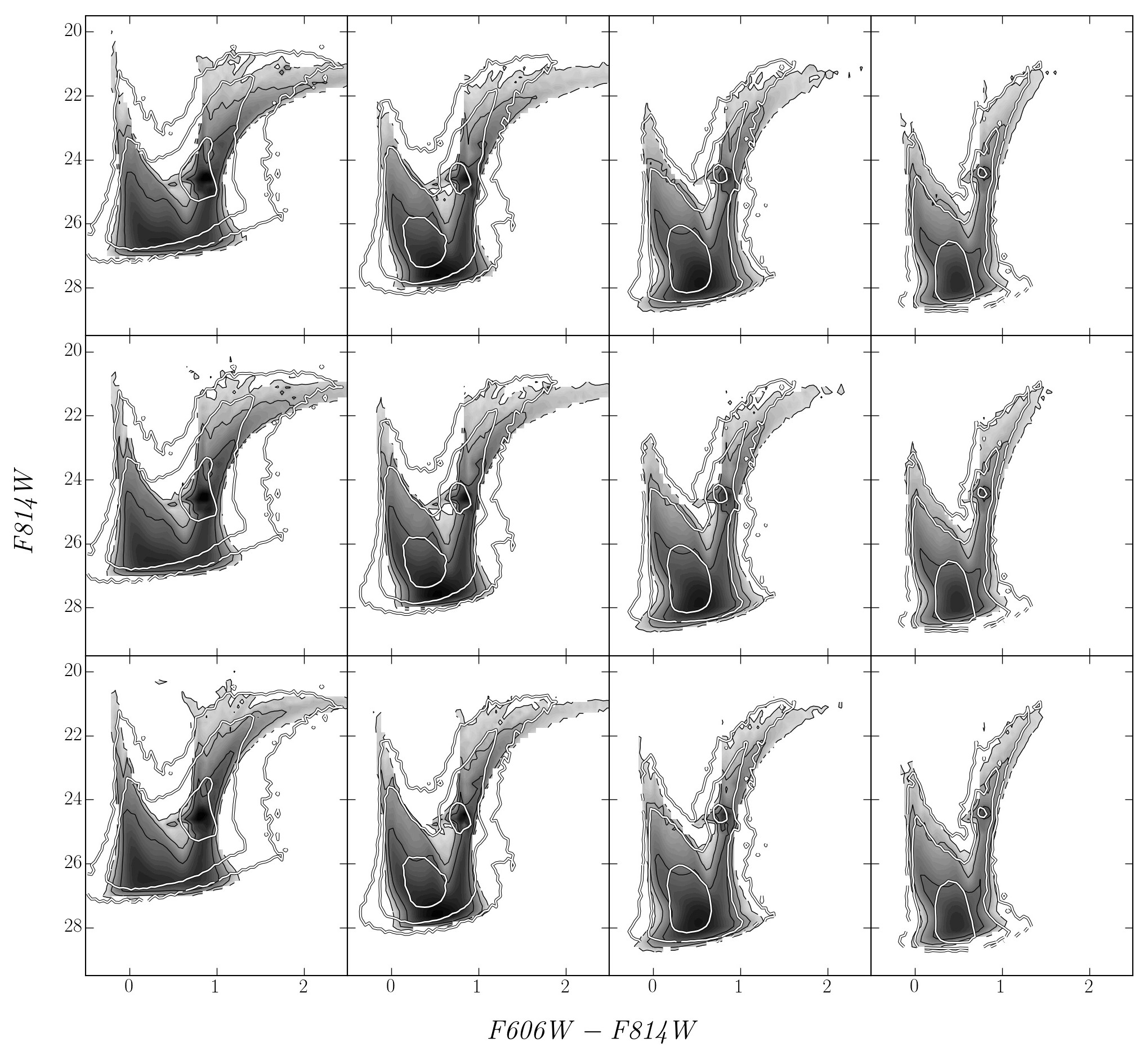,width=120mm,clip=}}
\vspace{1mm}
\captionb{4}
{The same as in Fig.\,2, but for the model C.}}
\end{figure}

The model C with the lowest value of $\epsilon$ is compared with the observations in
Fig.\,4. The low values of $\epsilon$ suppress gas outflow from the galaxy due to relatively
low-mass stellar populations forming in the disk. As a result of this effect, even
models with highest $\eta$ values shows too metallic stellar populations throughout
the galaxy disk.

\sectionb{5}{CONCLUSIONS}

We have demonstrated the capability of our improved stochastic 2-D models
of galaxy disk evolution to analyze resolved stellar photometry data of the galaxy
M33. The method of synthetic CMDs was proved to be a powerful tool to disentangle
 degeneracies of SF history in the cases when a turn-off of the main sequence
is below the detection limit of photometry. The synthetic CMDs help to break
parameter degeneracies discussed in our previous paper (Mineikis \& Vansevi{\v c}ius
2014a). Therefore, putting together our galaxy disk models, galaxy radial pro-
file observations performed in a wide range of wavelength -- from UV to radio,
as well as high precision and high resolution stellar photometry observations with
HST, we have demonstrated that degeneracies of the SF history parameters can
be resolved.

\thanks{We are thankful to Dr. Benjamin Williams for providing the HST/ACS WFC stellar photometry catalog of the galaxy M33. This research was partly funded by grant No. MIP-074/2013 from the Research Council
of Lithuania.}

\References

\refb Aparicio A., Hidalgo S.~L.\ 2009, AJ, 138, 558
\refb Asplund M., Grevesse N., Sauval A.~J., Scott P.\ 2009, ARA\&A, 47, 481
\refb Baumgartner V., Breitschwerdt D.\ 2013, A\&A, 557, AA140 
\refb Bressan A., Marigo P., Girardi L. et al.\ 2012, MNRAS, 427, 127 
\refb Cardelli J.~A., Clayton G.~C., Mathis J.~S.\ 1989, ApJ, 345, 245
\refb Chen Y., Girardi L., Bressan A. et al.\ 2014, MNRAS, 444, 2525 
\refb Corbelli E., Salucci P.\ 2000, MNRAS, 311, 441
\refb Fakhouri O., Ma C.-P., Boylan-Kolchin M.\ 2010, MNRAS, 406, 2267
\refb Ferguson A., Irwin M., Chapman S. et al.\ 2007, \textit{Island Universes -- Structure and Evolution of Disk Galaxies}, Springer, p. 239
\refb Gordon K.~D., Clayton G.~C., Misselt K.~A., Landolt A.~U., Wolff, M.~J.\ 2003, ApJ, 594, 279
\refb Harris J., Zaritsky D.\ 2009, AJ, 138, 1243 
\refb Heyer M.~H., Corbelli E., Schneider S.~E., Young J.~S.\ 2004, ApJ, 602, 723
\refb Kroupa P.\ 2002, Science, 295, 82 
\refb Lewis A.~R., Dolphin A.~E., Dalcanton J.~J. et al.\ 2015, ApJ, 805, 183
\refb Mineikis T., Vansevi{\v c}ius V.\ 2014a, Baltic Astronomy, 23, 209
\refb Mineikis T., Vansevi{\v c}ius V.\ 2014b, Baltic Astronomy, 23, 221
\refb Mu{\~n}oz-Mateos J.~C., Gil de Paz A., Boissier S. et al.\ 2007, ApJ, 658, 1006
\refb Rosolowsky E., Simon J.~D.\ 2008, ApJ, 675, 1213 
\refb Rubele S., Girardi L., Kerber L. et al.\ 2015, MNRAS, 449, 639
\refb Tang J., Bressan A., Rosenfield P. et al.\ 2014, MNRAS, 445, 4287
\refb Williams B. F., Dalcanton J.~J., Dolphin A.~E., Holtzman J., Sarajedini A.\ 2009, ApJL, 695, L15 

\end{document}